\documentclass[twocolumn, amssymb, aps, prl, superscriptaddress, floatfix]
{revtex4-2}

\usepackage{dcolumn}
\usepackage{bm}
\usepackage{amsmath}
\usepackage{calc}
\usepackage{xcolor}
\usepackage{multirow}

\usepackage{mathtools}
\usepackage{amsthm}

\usepackage[normalem]{ulem}

\usepackage{soul}

\usepackage[most]{tcolorbox}

\usepackage[colorlinks=true, allcolors=blue]{hyperref}
\usepackage[makeroom]{cancel}

\usepackage{listings}
\usepackage{color}

\usepackage{float}

\definecolor{dkgreen}{rgb}{0,0.6,0}
\definecolor{gray}{rgb}{0.5,0.5,0.5}
\definecolor{mauve}{rgb}{0.58,0,0.82}
\definecolor{orange}{rgb}{1.0, 0.5, 0.0}

\newcommand{\DA}[1]{\textcolor{blue}{#1}}

\begin{document}
\title{Superconductivity induced by altermagnetic spin fluctuations
in high-pressure MnB$_4$}

\author{Danylo Radevych}
\email{Corresponding author: dradevyc@gmu.edu}
\affiliation{Department of Physics and Astronomy,
George Mason University, Fairfax, VA 22030, USA}

\author{Merc\`{e} Roig}
\email[]{roigserv@uwm.edu}
\affiliation{Department of Physics,
University of Wisconsin-Milwaukee,
Milwaukee, WI 53201, USA}

\author{Daniel F. Agterberg}
\email[]{agterber@uwm.edu}
\affiliation{Department of Physics,
University of Wisconsin-Milwaukee,
Milwaukee, WI 53201, USA}

\author{Igor I. Mazin}
\email{imazin2@gmu.edu}
\affiliation{Department of Physics and Astronomy,
George Mason University, Fairfax, VA 22030, USA}
\affiliation{Quantum Science and Engineering Center, George Mason University,
Fairfax, VA 22030, USA}


\begin{abstract}
Recent experiments found superconductivity in nonmagnetic
MnB$_4$ with a high critical temperature ($T_{\text{c}}$) reaching
14~K at 158~GPa. However, \textit{ab initio} calculations of the
electron-phonon coupling predict a $T_{\text{c}}$ below 1~K,
suggesting that a conventional mechanism cannot explain this
phenomenon.
In this Letter, we find that MnB$_4$ is close
to an altermagnetic instability in density-functional
theory calculations.
We propose that the superconductivity
is driven by altermagnetic spin fluctuations.
To verify the pairing symmetry, we have constructed a
two-orbital tight-binding model, where boron states at the Fermi level
are integrated out. Using this model, we identify an extended-$s$ symmetry
as the leading pairing
instability.
If confirmed, this will be
the first reported case of superconductivity
driven by altermagnetic spin fluctuations.
\end{abstract}

\maketitle



\paragraph{Introduction ---}

\begin{figure}[t]
\includegraphics[width=1\columnwidth]{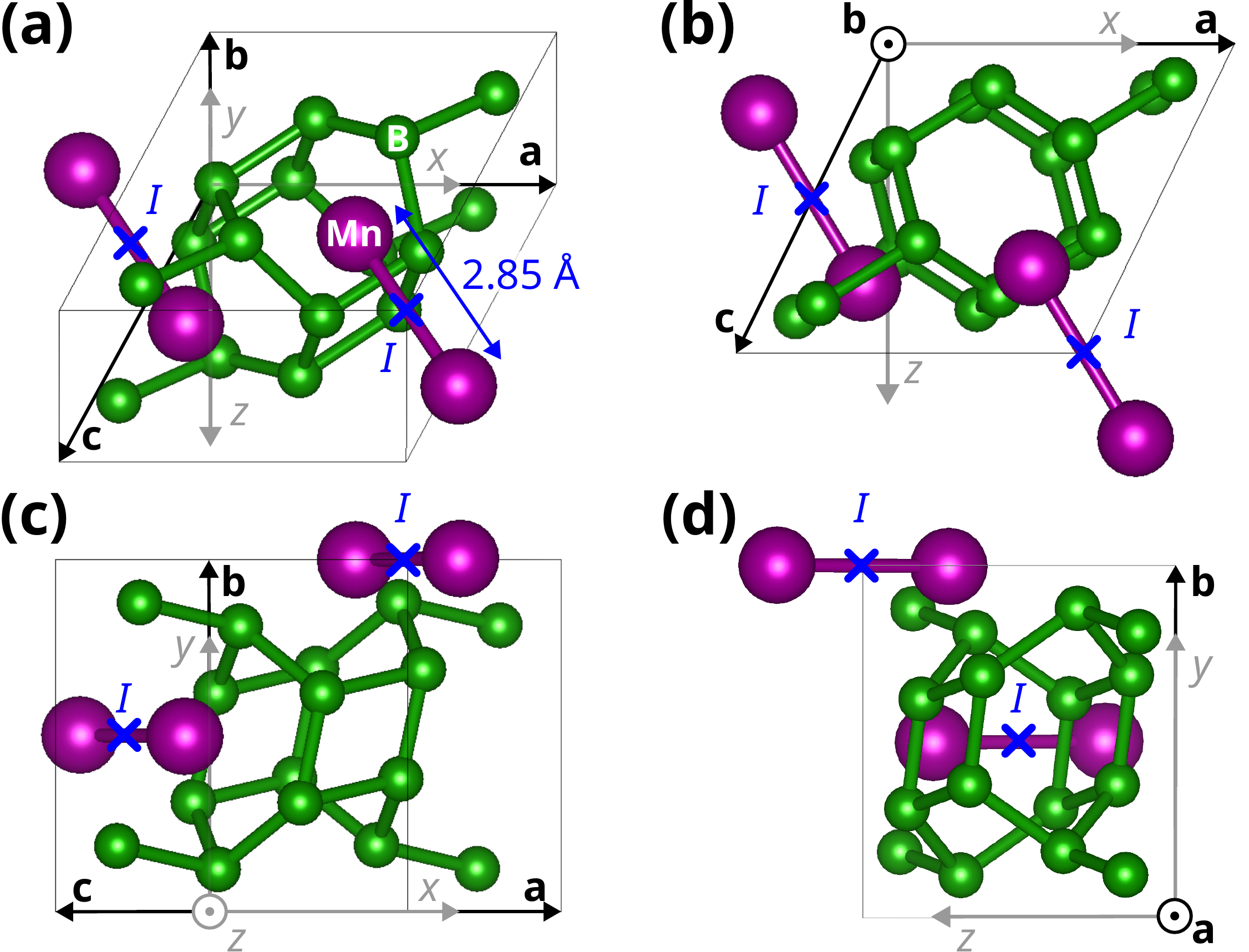}
\caption{$P2_1/c$ MnB$_4$ primitive cell at 158~GPa viewed from
several directions: (a) perspective view, (b) along the $y$-axis, (c)
along the $z$-axis, and (d) along the $x$-axis. Cartesian axes: gray
arrows; non-orthogonal lattice vectors ${\bf a}$, ${\bf b}$, and
${\bf c}$: black arrows; Mn: purple; B: green; Mn--Mn dimer and
inversion ($I$) centers: blue crosses.}
\label{fig:primcell}
\end{figure}

Following the introduction of altermagnets as a new class
of materials
\cite{Smejkal2022i, Smejkal2022ii},
the question arose whether altermagnetism could drive
superconductivity
\cite{Mazin2025, *Mazin2022}.
Altermagnetic (AM) spin fluctuations are of particular
interest since they can persist in phases lacking
long-range magnetic order, as such order is typically
detrimental to superconductivity \cite{Abrikosov1960}.
Confirmed cases of superconductors with antiferromagnetic
(AFM) spin fluctuations---characterized by the local
magnetic ordering vector ${\bf q} \neq {\bf 0}$ and
local net magnetization ${\bf M} = {\bf 0}$---have been
reported \cite{Sigrist1991, Moriya2003, Lee2006, Mazin2024}.
Similarly,
materials exhibiting
ferromagnetic (FM) spin fluctuations with
${\bf q} = {\bf 0}$ and ${\bf M} \neq {\bf 0}$
have also been suggested
\cite{Mazin1997, Hirschfeld2011, Hosono2015, Mazin2024}.
Although a recent theoretical study \cite{Wu2025}
presented a superconducting phase diagram for the remaining
possibility of AM spin fluctuations (${\bf q} = {\bf 0}$ and
${\bf M} = {\bf 0}$) that induce magnetic moments on two sites
within the same primitive cell \DA{\cite{Roig2024,Antonenko2025}}, no specific material utilizing this
mechanism has yet been identified.

Recent resistivity measurements \cite{Xiang2024} in nonmagnetic (NM)
$P2_1/c$ MnB$_4$ [see Figs.~\ref{fig:primcell}(a)--\ref{fig:primcell}(d)
and Secs.~I and V of the Supplemental Material (SM)~\cite{SM} for
crystallographic details] \cite{Fruchart1960, Andersson1969,
Andersson1970, Kolmogorov2010, Wang2011, Bialon2011, Gou2012,
Niu2012, Gou2013, Gou2014, Knappschneider2014, VanDerGeest2014,
Liang2015, Bykova2015, Steinki2017, Hajinazar2021} demonstrated the
emergence of superconductivity at pressures above 30~GPa, with a
critical temperature ($T_{\text{c}}$) of 14.2~K at
158~GPa. Since conventional electron-phonon coupling (EPC)
calculations based on density-functional perturbation theory (DFPT)
predicted $T_{\text{c}} < 1$~K, the superconductivity was speculated
to originate from an unconventional pairing mechanism associated with
Mn magnetic moments.
However,
the specific spin fluctuations driving the superconductivity have not
yet been identified.

In this Letter, we demonstrate that
spin fluctuations are, indeed, present in this system,
and their altermagnetic symmetry
leads to an unconventional extended-$s$ superconductivity.
This strongly suggests that MnB$_4$ represents the first
known material where superconductivity originates from proximity to altermagnetism.
Our density-functional theory (DFT) results
supporting this picture are described in detail below, as is the
effective minimal DFT-based
two-orbital tight-binding (TB) model.
The corresponding analysis of the superconducting
order parameter is also discussed.
Additionally, the DFPT EPC calculations performed in this work
are reported in Sec.~III of the SM \cite{SM}.

\paragraph{Conventional mechanism ---}

\begin{figure}[tbph]
\includegraphics[width=0.8\columnwidth]{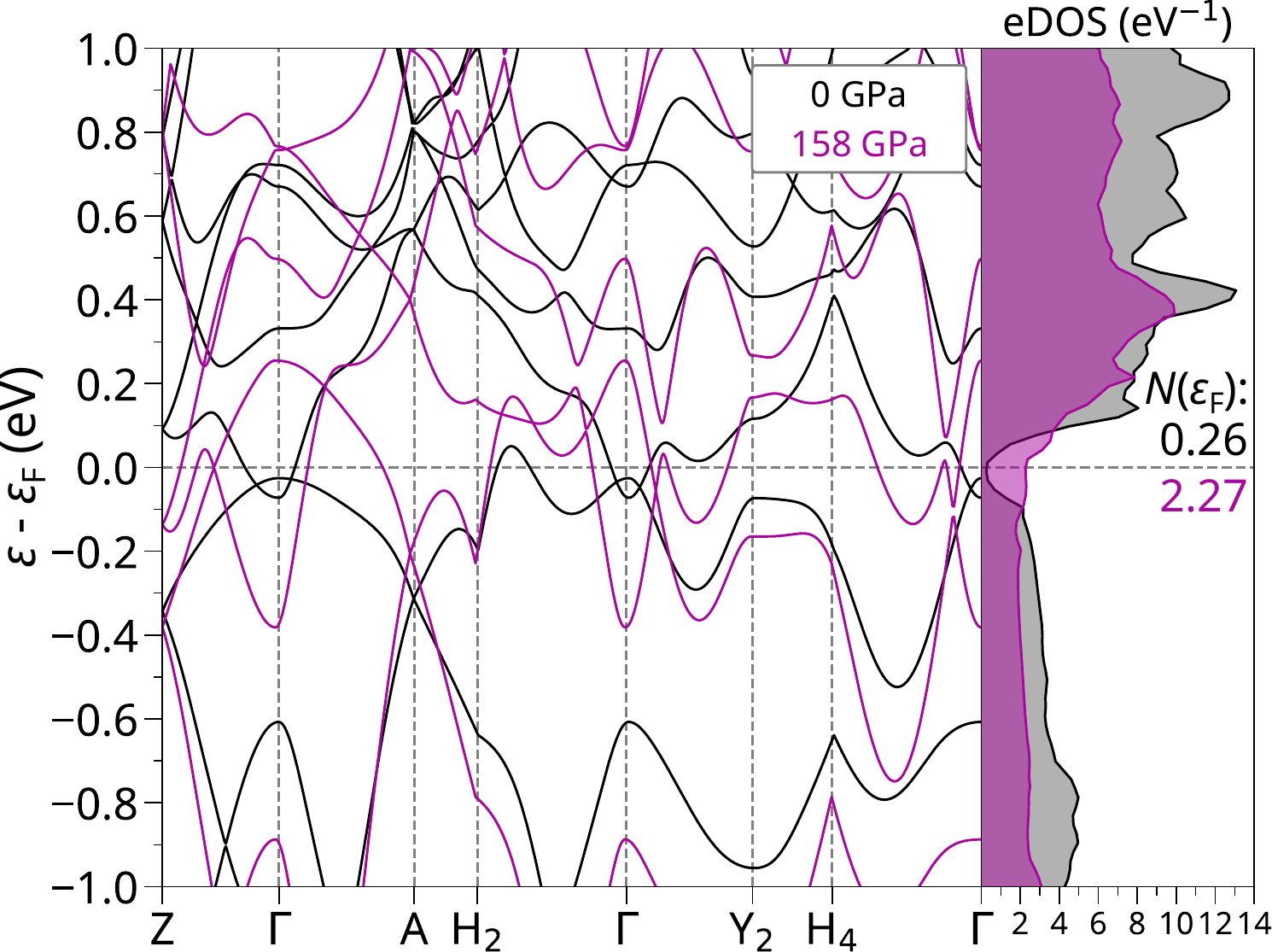}
\caption{Electronic band structure and density of states
(eDOS) of nonmagnetic $P2_1/c$ MnB$_4$ at 0~GPa (black) and 158~GPa
(purple). The eDOS is shown for both spin channels, with
corresponding values at the Fermi level $N(\varepsilon_{\text{F}})$
indicated near the horizontal dashed gray line.}
\label{fig:electrons}
\end{figure}

The electronic band structure and density of states (eDOS) of NM
MnB$_4$ at 0 and 158~GPa are shown in Fig.~\ref{fig:electrons}. Under
pressure, the semimetallic phase undergoes a transition to a metallic
state. In particular, the eDOS at the Fermi level
$N(\varepsilon_{\text{F}})$ increases from 0.07~eV$^{-1}$ per Mn at
0~GPa to 0.57~eV$^{-1}$ per Mn at 158~GPa.
Despite a notably higher
$N(\varepsilon_{\text{F}})$, the
results presented in Sec.~III of the SM \cite{SM} show that the
conventional critical temperature of MnB$_4$ due to EPC is below
1~K even at 158~GPa --- significantly lower than the experimental value of 14~K.
This large discrepancy between the calculated and experimentally measured
values strongly suggests that superconductivity in MnB$_4$ is not
driven by electron-phonon coupling.

\paragraph{Spin fluctuations ---}

Superconductivity driven by spin fluctuations typically emerges near
a quantum critical point (QCP) induced by pressure or doping
\cite{Alireza2009, Wilson2010, Hirschfeld2011, Luo2015, Wu2025}.
The tendency of Mn
ions to exhibit strong local magnetic moments in almost all their
compounds \cite{Pauling1931, Zener1951, Goodenough1955, Kubler1983}
and
the absence of long-range magnetic order in MnB$_4$ at zero pressure
\cite{Knappschneider2014, Gou2014, Steinki2017}
suggest that the QCP in MnB$_4$ is located at a slightly negative pressure.
This is somewhat different from other spin-driven superconductors,
where superconductivity sets in right at the QCP, or even slightly before
\cite{Mathur1998, Shibauchi2014}.
Apparently, such behavior in MnB$_4$ is arrested by the anomalously low eDOS,
and superconductivity only emerges at a higher pressure,
when sufficient eDOS has been attained.

A standard method to investigate the underlying magnetic order
that is not realized in normal DFT calculations
is to apply a Hubbard $U$ correction (``DFT$+U$'')
to enhance the propensity for
magnetism \cite{Petukhov,Das2023}.
This correction is not necessarily physically justified,
but it provides a gauge of how
close MnB$_4$ is to a magnetic transition and what the hidden
magnetic order is.

\begin{figure}[tbhp]
\includegraphics[width=\columnwidth]{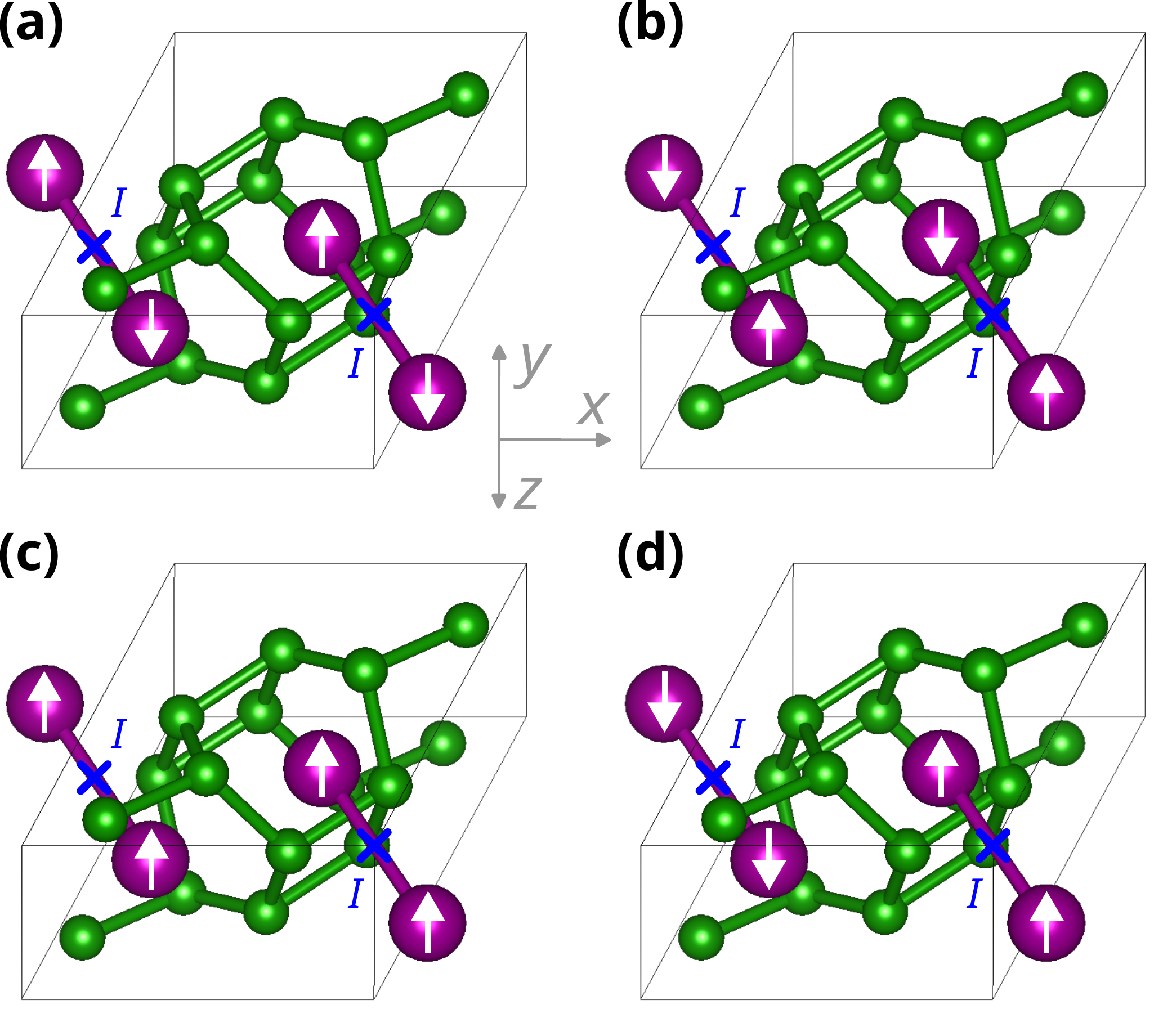}
\caption{Considered magnetic configurations of MnB$_4$: (a)
antiferromagnetic 1 (AFM~1), (b) antiferromagnetic 2 (AFM~2), (c)
ferromagnetic (FM), and (d) altermagnetic (AM). Primitive-cell edges:
black solid lines; Cartesian axes: gray arrows; Mn--Mn dimer and
inversion ($I$) centers: blue crosses; Mn magnetic moments: white
arrows.}
\label{fig:magconfig}
\end{figure}

\begin{figure}[tbhp]
\includegraphics[width=0.85\columnwidth]{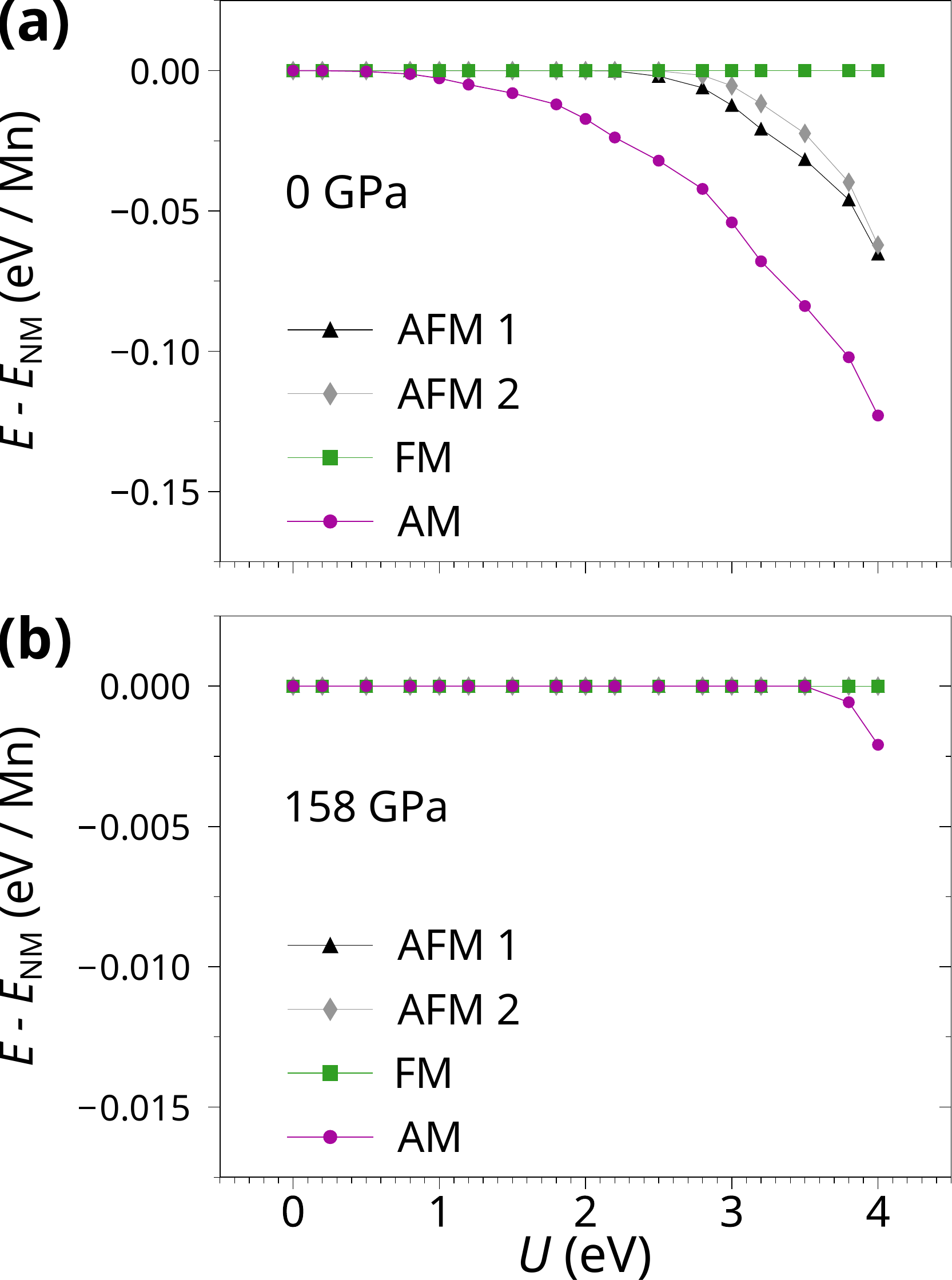}
\caption{Relative energies of magnetic $P2_1/c$ MnB$_4$
configurations with respect to the nonmagnetic (NM) state as a
function of the Hubbard $U$ parameter at (a)~0~GPa and (b)~158~GPa.
Abbreviations denote antiferromagnetic (AFM~1 and AFM~2),
ferromagnetic (FM), and altermagnetic (AM) orders. Note the
different energy scales used for 0 and 158~GPa.}
\label{fig:u}
\end{figure}

The $P2_1/c$ MnB$_4$ primitive cell contains two Mn--Mn
dimers whose centers coincide with the inversion $I$
centers
[Figs.~\ref{fig:primcell}(a)--\ref{fig:primcell}(d)].
The four distinct Mn sites allow for four
different magnetic states with ${\bf q} = {\bf 0}$.
These magnetic configurations within the primitive
cell---two AFM, one FM, and one AM---are shown in
Figs.~\ref{fig:magconfig}(a)--\ref{fig:magconfig}(d).
Their DFT$+U$-calculated total energies with respect to
the NM state for various values of $U$ at 0 and 158~GPa
are shown in Figs.~\ref{fig:u}(a) and \ref{fig:u}(b),
respectively.
The FM state is not stabilized at any pressure.
At 0~GPa, values of
$U \approx 1$~eV and $U \approx 2.5$~eV are required for the Mn atoms to
stabilize the AM and two AFM configurations, respectively.
The AM state
remains the lowest-energy phase for all higher values of $U$.
At 158~GPa, the AM configuration stabilizes at $U \approx 3.5$~eV
with an energy $\sim2.5$~meV~/~Mn lower than the NM state.
Since DFT$+U$ is a mean-field theory, these results do not
account for the suppression of the magnetic order by
spin fluctuations under pressure.
However, they demonstrate that the dominant
spin fluctuations in MnB$_4$ must be altermagnetic,
as the AM state consistently dominates the magnetic energy landscape.

\begin{figure}[t]
  \includegraphics[width=1\columnwidth]{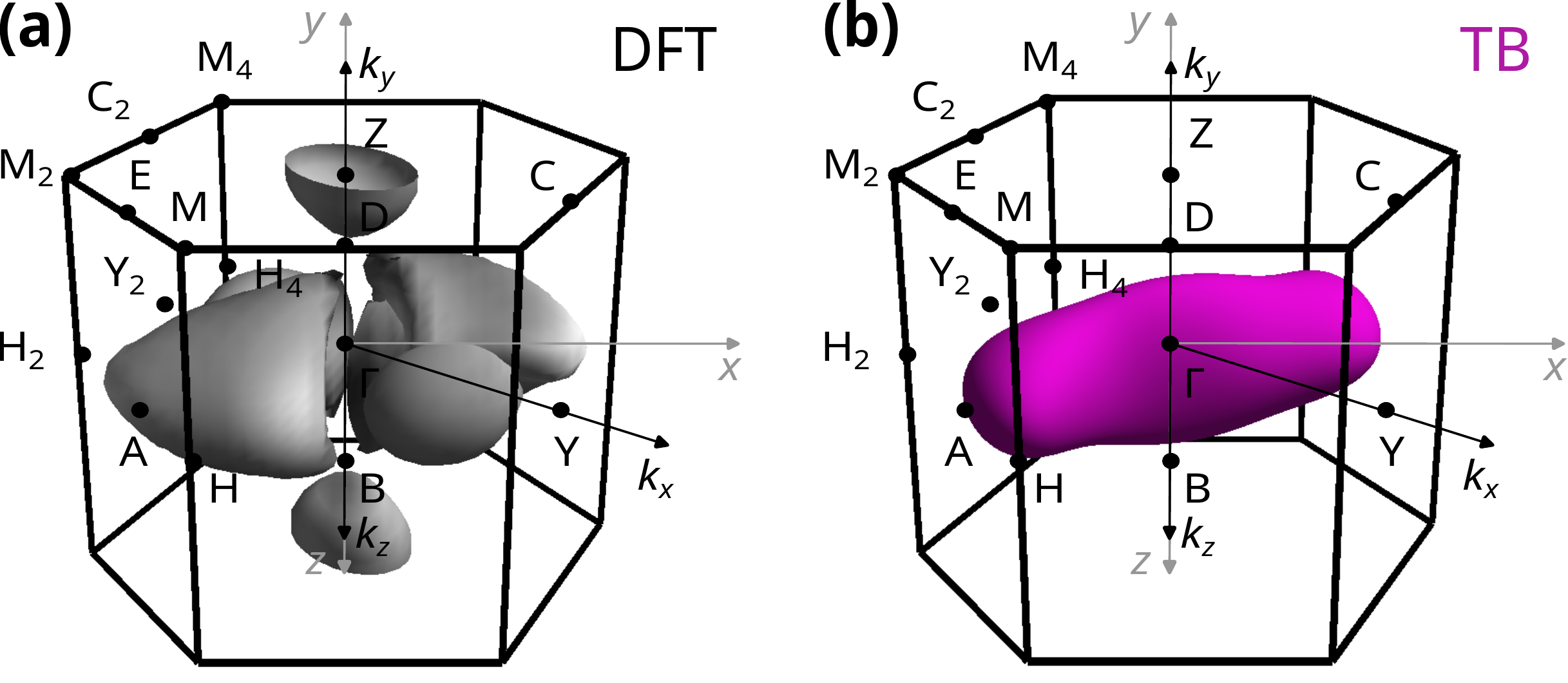}
  \caption{Comparison of the Fermi surfaces obtained from the (a)
  DFT-calculated electronic structure and (b) minimal tight-binding
  model.}
  \label{fig:tb_fs}
\end{figure}

\paragraph{Minimal tight-binding model ---}
The electronic structure of MnB$_4$ at 158~GPa
[Fig.~\ref{fig:electrons} and Fig.~S2(b) of the SM \cite{SM}]
consists of entangled B and Mn bands crossing the Fermi
level $\varepsilon_{\text{F}}$. The resulting DFT Fermi surface,
shown in Fig.~\ref{fig:tb_fs}(a), is complex due to the presence of B
pockets surrounding a central ``maggot''-shaped sheet with dominant Mn
character. Since spin fluctuations on Mn ions are expected to couple
with the Mn rather than B states, one needs to
exclude the B degrees of freedom and select only the relevant Mn
states near $\varepsilon_{\text{F}}$.
Furthermore, a close inspection reveals that the Mn states near
the Fermi surface arise from Mn--Mn dimers shown
in Figs.~\ref{fig:magconfig}(a)--\ref{fig:magconfig}(d).
Therefore, we replace each dimer with an effective ``bond'' orbital located
at its midpoint, which is an inversion center [blue crosses in
Figs.~\ref{fig:primcell}(a)--\ref{fig:primcell}(d)].
Note that in the AM arrangement,
both atoms of each dimer have the same spin polarization,
and the inversion centers do not destroy the AM state.

Thus, a minimal model consists of two bond orbitals per
primitive cell at the inversion centers, in agreement with a class of minimal models commonly used for altermagnets \cite{Roig2024,Antonenko2025,Wu2025}. For the $P2_1/c$
space group, the symmetry-constrained Hamiltonian is given by
\begin{equation}
H({\bf k}) = \left(\epsilon_0({\bf k}) - \mu\right) \tau_0 +
\epsilon_x({\bf k}) \tau_x + \epsilon_z({\bf k}) \tau_z,
\label{eq:hamiltonian}
\end{equation}
where $\epsilon_0({\bf k})$ and $\mu=\varepsilon_F$ denote the dispersion for the two bond orbitals and chemical
potential, respectively.
The $\tau_i$ are the $2\times2$ Pauli matrices acting in orbital
space. The ${\bf k}$-dependent functions $\epsilon_0({\bf k})$,
$\epsilon_x({\bf k})$, and $\epsilon_z({\bf k})$ are
\begin{equation}
\begin{aligned}
\epsilon_0({\bf k})
&=
t_x \cos k_x
+
t_y \cos k_y
+
t_z \cos k_z
\\
&
+
u_1 \sin k_x \sin k_z
+
u_2 \cos k_x \cos k_z
\\
&
+
u_3 \cos k_x \cos k_y
+
u_4 \cos k_y \cos k_z
\\
&
+
u_5 \sin k_x \cos k_y \sin k_z
+
u_6 \cos k_x \cos k_y \cos k_z,
\\ 
\epsilon_x({\bf k})
&=
t_1 \cos \frac{k_y}{2} \cos \frac{k_z}{2}
+
t_2 \sin k_x \cos \frac{k_y}{2} \sin \frac{k_z}{2},
\\ 
\epsilon_z({\bf k})
&=
t_3 \sin k_x \sin k_y
+
t_4 \sin k_y \sin k_z, \nonumber
\end{aligned}
\end{equation}
where $t_i$ and $u_j$ are hopping parameters. The
wavevector ${\bf k}$ is given in reciprocal crystal coordinates
multiplied by $2\pi$, with directions relative to the Brillouin zone
shown in Fig.~\ref{fig:tb_fs}(b).

Here, the Hamiltonian in Eq.~\eqref{eq:hamiltonian}
is fitted to two Wannierized Mn electronic bands near the Fermi level
$\varepsilon_{\text{F}}$, only one of which crosses
$\varepsilon_{\text{F}}$ (see Sec.~IV of the SM \cite{SM} for details
regarding the Wannierization procedure). Although the $t_i$
parameters are the most important in the TB model, the terms with $u_j$
parameters are also allowed by symmetry and improve the agreement
with the DFT results. All TB parameters extracted from the
Wannierization are reported in Table~\ref{tab:tbparams}. The
resulting TB Fermi surface is compared to the DFT Fermi surface in
Figs.~\ref{fig:tb_fs}(a)--\ref{fig:tb_fs}(b). The central Mn feature
in Fig.~\ref{fig:tb_fs}(b) is now isolated from B admixture and can be
analyzed for its superconducting properties.

\begin{table}[tb]
\caption{\label{tab:tbparams}
Numerical values of the parameters in the tight-binding model.}
\begin{tabular}{crcrcr}
\hline\hline
\noalign{\smallskip}
Parameter & Value & Parameter & Value & Parameter & Value\\
  & (eV) & & (eV) & & (eV)\\
\noalign{\smallskip}
\hline
\noalign{\smallskip}
$\mu$ & $-$0.829 & $t_1$ &    0.669 & $u_1$ & $-$0.169\\
$t_x$ &    0.028 & $t_2$ & $-$0.376 & $u_2$ &    0.234\\
$t_y$ &    0.151 & $t_3$ & $-$0.002 & $u_3$ & $-$0.197\\
$t_z$ &    0.060 & $t_4$ &    0.013 & $u_4$ & $-$0.112\\
      &          &       &          & $u_5$ &    0.119\\
      &          &       &          & $u_6$ &    0.103\\
\noalign{\smallskip}
\hline\hline
\end{tabular}
\end{table}

\paragraph{Leading pairing channel ---}
The derivation of the superconducting gap equation for isotropic
Heisenberg AM spin fluctuations is detailed in Sec.~II of the SM
\cite{SM}. Unlike Ising spin fluctuations, which involve only one
spin orientation and allow for $s$-, $p$-, and $d$-wave pairing states
in altermagnets \cite{Wu2025}, the isotropic model significantly
suppresses the $p$-wave channel \cite{SM}  due to the factor of three enhancement in the
singlet pairing strength (see the SM \cite{SM} for more details).
Consequently, only intra-unit cell spin-singlet extended $s$-wave [with $\Delta({\bf k})\approx \cos(k_y/2)\cos(k_z/2)$] and
$d$-wave [with $\Delta({\bf k})\approx \sin(k_y/2)\sin(k_z/2)$] channels compete.
As shown in Figs.~S1(e) and S1(f) of the SM
\cite{SM}, the $d$-wave state is less favorable
due to nodal lines in the order parameter
that suppress the gap magnitude over a substantial
portion of the Fermi surface.
The resulting pairing eigenvalue, determining the
instability temperature, for the $s$-wave channel ($\lambda_s$) is much
larger than that for the $d$-wave channel ($\lambda_d$).

\paragraph{Conclusions ---}
The possibility of conventional electron-phonon superconductivity in
MnB$_4$ is conclusively ruled out. Instead, we argue that the observed
superconductivity is driven by altermagnetic spin
fluctuations, detected by means of DFT$+U$ calculations. Using a minimal
two-orbital tight-binding model fitted to Wannierized DFT bands,
and integrating out the irrelevant boron states, an extended
$s$-wave pairing channel is identified as the leading
superconducting instability. These results suggest that MnB$_4$
represents the first realized example of superconductivity
originating from altermagnetism, thereby opening a new avenue for
magnetism-driven pairing states.

\begin{acknowledgments}
\paragraph{Acknowledgements ---}
The authors thank Elena~R.~Margine and Aleksey~N.~Kolmogorov for
helpful discussions. This work was supported by the National Science
Foundation (Grants Nos. DMR-2320074 and DMREF-2323857).
I.M. acknowledges support from
Army Research Office under Cooperative Agreement Number W911NF-22-2-0173.
D.F.A and M.R. acknowledge support from the Simons
Foundation grant SFI-MPS-NFS-00006741-02.
D.R. also
acknowledges the Texas Advanced Computing Center (TACC) at The
University of Texas at Austin \cite{TACC} (Allocation Nos.
TG-DMR180071 and DMR22004) and the San Diego Supercomputer Center
(SDSC) at the University of California San Diego \cite{SDSC}
(Allocation No. TG-DMR180071) for providing the computational
resources used in this work.

\end{acknowledgments}

\appendix
\section{End matter}

\paragraph{DFT ---}
Electronic structure calculations were performed using the Vienna
\textit{Ab initio} Simulation Package (VASP) version 6.5.1
\cite{Kresse1993, Kresse1994, Kresse1996a, Kresse1996b, Kresse1999}.
Standard projector augmented-wave (PAW) \cite{Blochl1994}
pseudopotentials with the generalized gradient approximation (GGA)
of Perdew, Burke, and Ernzerhof (PBE) \cite{Perdew1996} were employed.
The calculations used a plane-wave energy cutoff of 580~eV, a
Methfessel-Paxton smearing \cite{Methfessel1989} of 0.27~eV, and a
$12\times12\times12$ $\Gamma$-centered ${\bf k}$-point grid.

Additionally, the results were verified using the PWscf
package of Quantum ESPRESSO (QE) version 7.4.1
\cite{Giannozzi2009, Giannozzi2017,
Giannozzi2020, QE}
as a required step before the DFPT calculations of
the electron-phonon coupling.
In QE, optimized norm-conserving Vanderbilt
pseudopotentials (ONCVPSP)
\cite{Hamann2013, Hamann2017} with the PBE
functional---obtained from the PseudoDojo library
\cite{vanSetten2018, Garrity2014,
Lejaeghere2016}---were used with a wavefunction energy
cutoff of 90~Ry and an automatically generated
$12\times12\times12$ ${\bf k}$-point grid.
The electronic structures from VASP and QE
were in excellent agreement.

Crystal structures and Fermi surfaces were
visualized with VESTA \cite{Momma2011} and
FermiSurfer \cite{Kawamura2019}, respectively.

\paragraph{DFPT ---}
EPC calculations were performed in QE using the same
pseudopotentials and energy cutoffs as the initial DFT calculations.
Other technical details are reported in Sec.~III of the SM \cite{SM}.

\paragraph{DFT$+U$ ---}
Total energy calculations for various values of the Hubbard $U$
parameter were performed in VASP. FM, AFM, and AM configurations
were symmetry-classified using the \textsc{amcheck} program
\cite{AMCHECK, AMCHECK-r1.0}.

\paragraph{Wannierization ---}
The interface between VASP 6.5.1 and Wannier90 3.1.0
\cite{Pizzi2020, Marzari2012} was used for the
Wannierization process, with further technical details
provided in Sec.~IV of the SM \cite{SM}.

\bibliography{references}

\end{document}


\title{Supplemental Material\\
Superconductivity induced by altermagnetic spin fluctuations
in high-pressure MnB$_4$}

\author{Danylo Radevych}
\email{Corresponding author: dradevyc@gmu.edu}
\affiliation{Department of Physics and Astronomy,
George Mason University, Fairfax, VA 22030, USA}

\author{Merc\`{e} Roig}
\email[]{roigserv@uwm.edu}
\affiliation{Department of Physics,
University of Wisconsin-Milwaukee,
Milwaukee, WI 53201, USA}

\author{Daniel F. Agterberg}
\email[]{agterber@uwm.edu}
\affiliation{Department of Physics,
University of Wisconsin-Milwaukee,
Milwaukee, WI 53201, USA}

\author{Igor I. Mazin}
\email{imazin2@gmu.edu}
\affiliation{Department of Physics and Astronomy,
George Mason University, Fairfax, VA 22030, USA}
\affiliation{Quantum Science and Engineering Center, George Mason University,
Fairfax, VA 22030, USA}

\maketitle



\supplementarysection

\section{Crystallographic data}
Experimental lattice parameters for the $P2_1/c$ MnB$_4$ structure
at 0 and 158~GPa were obtained from Ref.~\cite{Xiang2024}.
Parameters were converted from the convention where the only
non-right angle, $\beta$, exceeds $120^\circ$ to the standard
convention ($\beta < 120^\circ$) using FINDSYM 7.1.7
\cite{FINDSYM, Stokes2005}. The standard lattice, illustrated in
Figs.~1(a)--1(d) of the main text, is derived by applying the
transformation ${\bf a}^\prime = {\bf a} + {\bf c}$,
${\bf b}^\prime = - {\bf b}$, and ${\bf c}^\prime = - {\bf c}$.
Crystallographic data are summarized in Table~\ref{tab:crysparam}.
All density-functional theory (DFT) calculations used experimental
lattice parameters with relaxed internal atomic positions.

\begin{table}[htbp]
\caption{\label{tab:crysparam}Crystallographic parameters of the
considered $P2_1/c$ MnB$_4$ structures.
Lattice parameters are obtained from Ref.~\cite{Xiang2024}.}
\begin{tabular}{ccccccccc}
\hline\hline
\noalign{\smallskip}
\multicolumn{1}{c}{Structure} &
Pressure &
$a, b, c$ &
$\alpha, \beta, \gamma$ &
Atom &
$x$, $y$, $z$ &
Wyckoff &
Mn--Mn links\\
&
(GPa) &
(\AA{}) &
($^\circ$) &
&
(fractional) &
position &
(\AA{})\\
\noalign{\smallskip}
\hline
\noalign{\smallskip}
MnB$_4$ & 0 & 5.48, 5.37, 5.50 & 90, 115, 90 &
Mn    & 0.27, 0.50, 0.77 & 4e & 2.70,~3.20\\
& & & &
B$_1$ & 0.13, 0.18, 0.87 & 4e\\
& & & &
B$_2$ & 0.16, 0.63, 0.34 & 4e\\
& & & &
B$_3$ & 0.33, 0.13, 0.68 & 4e\\
& & & &
B$_4$ & 0.36, 0.69, 0.14 & 4e\\
        & 158 & 4.97, 4.99, 4.93 & 90, 116, 90 &
Mn    & 0.27, 0.50, 0.77 & 4e & 2.40,~2.85\\
& & & &
B$_1$ & 0.12, 0.19, 0.88 & 4e\\
& & & &
B$_2$ & 0.17, 0.62, 0.34 & 4e\\
& & & &
B$_3$ & 0.32, 0.12, 0.69 & 4e\\
& & & &
B$_4$ & 0.37, 0.69, 0.14 & 4e\\
\noalign{\smallskip}
\hline\hline
\end{tabular}
\end{table}

\section{Derivation of the gap equation}

In this section, we use the minimal tight-binding model described in the main text to derive the gap equation and examine which pairing channel corresponds to the leading superconducting instability. To this end, we solve the linearized gap equation by projecting onto the different gap symmetries.

The general form of the tight-binding Hamiltonian by considering the two Mn--Mn dimers corresponds to
\begin{equation}
	H_0 (\kv) = (\epsilon_0(\kv) - \mu) \tau_0 + \epsilon_x(\kv) \tau_x + \epsilon_z(\kv) \tau_z,
\end{equation}
with the form of the momentum-dependent hoppings given in the main text. 
To include the coupling to altermagnetic fluctuations, we use an approach inspired by Ref.~\cite{Wu2025}. In this work, the key result was that altermagnetic fluctuations stabilized predominantly intra-unit cell spin-singlet and spin-triplet states. Provided we are not too close to the altermagnetic quantum critical point, the stability of these pairing states can be understood as originating from nearest-neighbor antiferromagnetic interactions  \cite{Wu2025}. For physical transparency, we explicitly begin with such an effective interaction and justify this by noting that it reproduces the intra-unit pairing states found in Ref.~\cite{Wu2025}.  This also allows us to highlight the key difference between our approach and the approach used in Ref.~\cite{Wu2025}. Specifically, we consider a spin-rotational invariant, or Heisenberg, interaction while an Ising-like interaction was considered in Ref.~\cite{Wu2025}. These two interactions are:
\begin{equation}
V_{\rm Heisenberg}=J\sum_{\langle i,j \rangle} \vec{S}_{1,i}\cdot \vec{S}_{2,j}
\end{equation}
and
\begin{equation}
V_{\rm Ising}=J\sum_{\langle i,j \rangle} S^z_{1,i}\cdot S^z_{2,j}
\end{equation}
where $\langle i,j \rangle$ denotes nearest neighbors and $\vec{S}_{\alpha,j}$ ($S^z_{\alpha,j}$) denotes the spin ($z$-component of spin) on site $\alpha$ in unit cell $j$. Decomposing these interactions in the pairing channel yields the effective pairing interactions
\begin{equation}
    H_{\rm Heisenberg} =  \sum_{\kv,\kv',s,s'} V_{\kv,\kv'} c^\dagger_{\kv 1s} c^\dagger_{-\kv 2s'} c_{-\kv' 2 s'} c_{\kv' 1 s},
    \label{eq:int_Heisenberg}
\end{equation}
and
\begin{equation}
    H_{\rm Ising} =  \sum_{\kv,\kv',s,s'} V_{\kv,\kv'} (\delta_{s,s'}-\delta_{s,-s'})c^\dagger_{\kv 1s} c^\dagger_{-\kv 2s'} c_{-\kv' 2 s'} c_{\kv' 1 s},
    \label{eq:int_Ising}
\end{equation}
where $s,s'$ denote the sum over spin and $V_{\kv,\kv'}= -J\cos(\frac{k_y-k_y'}{2}) \cos(\frac{k_z-k_z'}{2})$ is the nearest-neighbor interaction.

In general, these interactions allow both spin-singlet and spin-triplet solutions. For $V_{\rm Heisenberg}$, the spin-singlet solution is attractive with an interaction $-3J$, while the spin-triplet solutions are repulsive with an interaction $J$. However, for $V_{\rm Ising}$ the spin-singlet solution is attractive with an interaction $-J$, while the three spin-triplet solutions have different interactions. Denoting the spin structure of these solutions as $\Delta_i=\sigma_i(i\sigma_y)$, we find that $\Delta_z$ is attractive with an interaction $-J$, while $\Delta_x$ and $\Delta_y$ are repulsive with an interaction $J$. Hence, $V_{\rm Ising}$ allows both spin-singlet and spin-triplet solutions as found in Ref.~\cite{Wu2025}, while $V_{\rm Heisenberg}$ only allows spin-singlet solutions.

We now restrict our attention to $V_{\rm Heisenberg}$ and consider only the stable spin-singlet solutions. 
Projecting the nearest-neighbor interaction into the different symmetry channels, we obtain the following two contributions to the singlet channel:
\begin{equation}
	V^{\rm sing}_{\kv,\kv'} = \, - J \Bigg( \cos\frac{k_y}{2} \cos\frac{k_z}{2} \cos\frac{k_y'}{2} \cos\frac{k_z'}{2} + \sin\frac{k_y}{2} \sin\frac{k_z}{2} \sin\frac{k_y'}{2} \sin\frac{k_z'}{2} \Bigg)
    \label{eq:singlet_int}
\end{equation}
Note that the interaction in Eq.~\eqref{eq:int_Heisenberg} is between different sublattices, and therefore this reveals two possible pairing states: a $d$-wave gap $\tau_x \sin\frac{k_y}{2} \sin\frac{k_z}{2} (i\sigma_y)$ and an extended $s$-wave gap $\tau_x \cos\frac{k_y}{2} \cos\frac{k_z}{2} (i\sigma_y)$.

The gap equation can be written as~\cite{Wu2025}
\begin{equation}
	\Delta_i(\kv)=\sum_{\kv^\prime,j} V_{\kv,\kv^\prime} \Gamma^{i,j}_{\kv^\prime} \Delta_j (\kv^\prime),
    \label{eq:gap_equation}
\end{equation}
where $\Gamma^{i,j}_\kv=-T\sum_{i\omega_n}\Tr[\hat{G}_h \tau_i \hat{G}_p \tau_j]$ is the projector onto the pairing channels, with $\hat{G}_p(\kv) = [i\omega_n - H_0(\kv)]^{-1}$ and $\hat{G}_h(\kv) = [i\omega_n + H_0^T(\kv)]^{-1}$ corresponding to the bare particle and hole Green's functions.
To determine the form of the projector, we write the particle and hole Green's function in the band basis,
\begin{align}
& \hat{G}_p = \sum_{a=\pm} G_{0,p}^a(\kv,i\omega_n) P_\kv^a, \ \textrm{with} \  G_{0,p}^{\pm}(\kv,i\omega_n) = \frac{1}{i\omega_n - E_{\kv}^{\pm}}, \\
& \hat{G}_h = \sum_{a=\pm} G_{0,h}^a(\kv,i\omega_n) P_\kv^a, \ \textrm{with} \ G_{0,h}^{\pm}(\kv,i\omega_n) = \frac{1}{i\omega_n + E_{\kv}^{\pm}},
\end{align}
where the dispersion for the minimal tight-binding model is $E_{\kv}^{\pm}=\epsilon_0 (\kv)\pm \sqrt{\epsilon_x^2(\kv) + \epsilon_z^2(\kv)}$. The projector for the particle and hole Green's function is the same since $H_0^T(-\kv)=H_0(\kv)$, and it can be written as
\begin{equation}
    P_{\kv}^a = \frac{1}{2}\Bigg(\mathds{1} + a \frac{\epsilon_x(\kv) \tau_x + \epsilon_z(\kv) \tau_z}{\sqrt{\epsilon_x^2(\kv) + \epsilon_z^2(\kv)}}\Bigg),
\end{equation}
with $a=\pm$ denoting the two bands.

Since the $d$-wave and $s$-wave gaps are in the $\tau_x$ channel, we focus on the projector for these pairing states, which corresponds to
\begin{equation}
    \Gamma^{x,x}_\kv =-T\sum_{i\omega_n}\Tr[\hat{G}_h \tau_x \hat{G}_p \tau_x] =- T \frac{2}{\epsilon_x^2(\kv) + \epsilon_z^2(\kv)} \Big[ \epsilon_z^2(\kv)(G_{0,p}^+ G_{0,h}^- + G_{0,p}^- G_{0,h}^+) + \epsilon_x^2 (\kv)(G_{0,p}^+ G_{0,h}^+ + G_{0,p}^- G_{0,h}^-) \Big].
\end{equation}
The DFT band structure in the main text reveals that only $E_{\kv}^+$ crosses the Fermi level, while $E_{\kv}^{-}$ is far below the chemical potential. Hence, we keep only the projection to the intraband states and focus on the $E_{\kv}^{+}$ band. In this case, the previous expression can be written as
\begin{equation}
    \Gamma^{x,x}_\kv = - T \frac{2\epsilon_x^2(\kv)}{\epsilon_x^2(\kv) + \epsilon_z^2(\kv)} \sum_{i\omega_n} \frac{1}{i\omega_n + E_{\kv}^{+}} \frac{1}{i\omega_n - E_{\kv}^{+}}
     = \frac{\epsilon_{x}^2(\kv)}{\epsilon_{x}^2(\kv) + \epsilon_{z}^2(\kv)} \frac{\tanh \left(\frac{E_{\kv}^{+}}{2T}\right)}{E_{\kv}^{+}}.
\end{equation}
Therefore, the gap equation in Eq.~\eqref{eq:gap_equation} is given by
\begin{equation}
    \Delta_x(\kv) = \sum_{\kv',\alpha={\{d,s\}}} - J\frac{\epsilon_{x}^2(\kv')}{\epsilon_{x}^2(\kv') + \epsilon_{z}^2(\kv')} g_{\alpha}(\kv) g_\alpha (\kv') \frac{\tanh \left(\frac{E_{\kv'}^{+}}{2T}\right)}{E_{\kv'}^{+}} \Delta_x (\kv'),
\end{equation}
with $g_{\alpha}(\kv)$ the basis functions for the $s$-wave and $d$-wave gap in Eq.~\eqref{eq:singlet_int}, corresponding to $g_s (\kv) = \cos\frac{k_y}{2} \cos\frac{k_z}{2}$ and $g_d (\kv) = \sin\frac{k_y}{2} \sin\frac{k_z}{2}$, respectively.

To determine the leading pairing channel, we solve the linearized gap equation on the Fermi surface as an eigenvalue problem.
In the presence of nearest-neighbor interactions only, we determine whether the leading superconducting instability corresponds to the $d$-wave or the extended $s$-wave state by projecting the linearized gap equation to the previous basis functions $g_{\alpha}(\kv)$,
\begin{equation}
    \lambda_\alpha = \frac{ - \int \frac{d \kv'}{v_{\kv'}}\int \frac{d \kv}{v_{\kv}} g_{\alpha}(\kv') V^{\rm s}(\kv,\kv') g_{\alpha}(\kv)}{\int \frac{d \kv}{v_{\kv}} g_{\alpha}^2(\kv')},
\end{equation}
with $V^{\rm s}(\kv,\kv')= -J\frac{\epsilon_x^2(\kv)}{\epsilon_x^2(\kv)+\epsilon_z^2(\kv)} \sum_{\alpha'} g_{\alpha'}(\kv) g_{\alpha'} (\kv')$ and $v_{\kv}$ the Fermi velocity magnitude. Hence, the eigenvalues for the $d$-wave and extended $s$-wave channels correspond to
\begin{align}
    \lambda_d &=   J \int d \kv
    \underbrace{
    \frac{1}{v_{\kv}} \frac{\epsilon_{x}(\kv)^2}{\epsilon_{x}(\kv)^2+\epsilon_{z}(\kv)^2}
    \overbrace{\Bigg(\sin\frac{k_y}{2} \sin\frac{k_z}{2}\Bigg)^2}^{g^2_{d}(\kv)}
    }_{\tilde{\lambda}_{d}(\kv)},
    \label{eq:lambda_d}\\
    \lambda_s &=   J \int d \kv
    \underbrace{
    \frac{1}{v_{\kv}} \frac{\epsilon_{x}^2(\kv)}{\epsilon_{x}^2(\kv)+\epsilon_{z}^2(\kv)}
    \overbrace{\Bigg( \cos\frac{k_y}{2} \cos\frac{k_z}{2}\Bigg)^2}^{g^2_{s}(\kv)}
    }_{\tilde{\lambda}_{s}(\kv)}.
    \label{eq:lambda_s}
\end{align}
The largest eigenvalue corresponds to the leading pairing instability.

\begin{figure}[t]
  \includegraphics[width=0.9\textwidth]{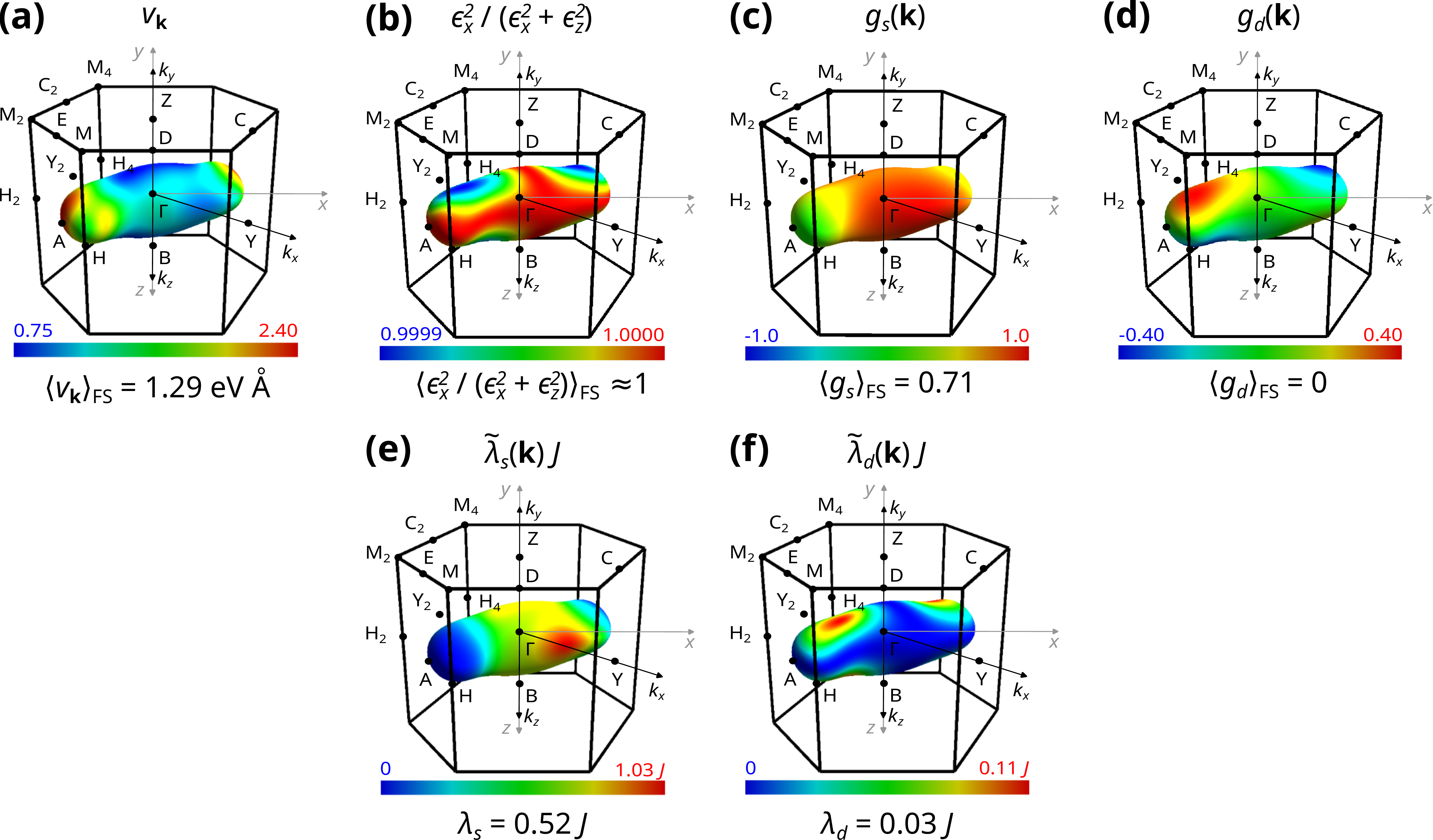}
  \caption{
  Projections on the Fermi surface of
  (a)~velocity magnitude $v_{\bf k} = \left|{\bf v}_{\bf k}\right|$,
  (b)~weighting factor $\epsilon_x^2({\bf k}) /
  \left(\epsilon_x^2({\bf k}) + \epsilon_z^2({\bf k})\right)$,
  (c)~pairing basis function $g_{s}({\bf k})$,
  (d)~pairing basis function $g_{d}({\bf k})$,
  (e)~pairing eigenvalue contribution $\tilde{\lambda}_s({\bf k})$,
  and (f)~pairing eigenvalue contribution $\tilde{\lambda}_d({\bf k})$.
  The corresponding values averaged over the Fermi surface are
  specified at the bottom of each plot, with $J$
  taken
  as a constant. Note that the distribution in (b) is
  nearly equal to unity and can thus be ignored.
  }
  \label{fig:fs_proj}
\end{figure}

In Figs.~\ref{fig:fs_proj}(a)--\ref{fig:fs_proj}(f) we show the projections on the Fermi surface of the terms entering Eqs.~\eqref{eq:lambda_d}--\eqref{eq:lambda_s}. As seen, the Fermi velocity is largest at the Brillouin zone edges, while the momentum dependence of the factor $\epsilon_{x}^2(\kv)/(\epsilon_{x}^2(\kv)+\epsilon_{z}^2(\kv))$ is almost negligible. The momentum dependence of the extended $s$-wave and $d$-wave pairing gaps is shown in Figs.~\ref{fig:fs_proj}(e) and \ref{fig:fs_proj}(f), respectively. Importantly, the $d$-wave gap has more nodes than the extended $s$-wave gap. As a consequence, comparing the eigenvalues for the two pairing channels (see Figs.~\ref{fig:fs_proj}(e)--\ref{fig:fs_proj}(f)),
we find that $\lambda_s$ is larger than $\lambda_d$
by more than an order of magnitude. Therefore, this reveals that the $s$-wave state is the leading pairing instability.

One may ask what the role of the Coulomb repulsion is, as it often favors a $d$-wave state. First of all, since our effective orbitals are actually dimer states, the bare on-site repulsion is greatly reduced. Second, even though we cannot evaluate the full frequency dependence of the pairing interaction, proximity to magnetism suggests that the relevant spin fluctuations are reasonably soft and the Tolmachev-Anderson-Morel logarithmic renormalization reduces the Coulomb pseudopotential to its typical values, i.e., considerably smaller than $\lambda$. With this in mind, we conclude that the Coulomb repulsion probably reduces the gigantic advantage of the $s$-state derived above, but falls quite short of stabilizing the $d$-state. In fact, we tested the (quite unphysical) limit of an infinitely strong on-site repulsion, and found that even then the $s$ state is still more favorable than the $d$ one.

\section{Electron-phonon coupling with and without doping}

Calculations of phonon modes and electron-phonon coupling (EPC) within the
density-functional perturbation theory (DFPT) framework
were performed using the Quantum ESPRESSO (QE) PHonon package
with the same
pseudopotentials, wavefunction energy cutoff,
and Brillouin zone sampling used for the
electronic structure calculations with QE.
A 4$\times$4$\times$4 phonon ${\bf q}$-point grid was employed,
and the ``crystal'' acoustic sum rule was applied to the interatomic
force constants.

In DFPT, the EPC constant $\lambda_{\text{EPC}}$ is calculated as \cite{Wierzbowska2006}
\begin{equation}
\begin{aligned}
\lambda_{\text{EPC}} &= 2 \int \frac{\alpha^2 F(\omega)}{\omega} d \omega
=
\frac{2}{N(\varepsilon_{\text{F}})} \sum_{{\bf q}\nu}
\frac{1}{\omega_{{\bf q}\nu}}
\sum_{mn} \sum_{{\bf k}}
\left| g_{mn\nu}({\bf k}, {\bf q}) \right|^2
\delta\left(\varepsilon_{m, {\bf k}+{\bf q}} - \varepsilon_{\text{F}}\right)
\delta\left(\varepsilon_{n, {\bf k}} - \varepsilon_{\text{F}}\right),
\end{aligned}
\label{eq:lambda}
\end{equation}
where $\omega$ is the phonon energy, $\alpha^2 F(\omega)$ is the
Eliashberg spectral function \cite{Eliashberg1960},
$N(\varepsilon_{\text{F}})$ is the
electronic density of states (eDOS)
at the Fermi energy $\varepsilon_{\text{F}}$,
$\omega_{{\bf q}\nu}$ is the phonon energy corresponding to mode
$\nu$ at wavevector ${\bf q}$,
$\varepsilon_{m, {\bf k}+{\bf q}}$ is the energy
of electronic state $m$ at the wavevector ${\bf k}+{\bf q}$,
and $g_{mn\nu}({\bf k}, {\bf q})$ is the EPC matrix element.
The Eliashberg spectral function is expressed in terms of the EPC matrix
elements and three Dirac delta-functions:

\begin{equation}
\begin{aligned}
\alpha^2 F(\omega) &=
\frac{1}{N(\varepsilon_{\text{F}})}
\sum_{mn} \sum_{{\bf q} \nu}
\delta\left(\omega - \omega_{{\bf q}\nu}\right)
\sum_{{\bf k}}
\left| g_{mn\nu}({\bf k}, {\bf q}) \right|^2
\delta\left(\varepsilon_{m, {\bf k}+{\bf q}} - \varepsilon_{\text{F}}\right)
\delta\left(\varepsilon_{n, {\bf k}} - \varepsilon_{\text{F}}\right).
\end{aligned}
\label{eq:a2f}
\end{equation}
In the presented calculations,
the EPC matrix elements were
interpolated to a 24$\times$24$\times$24 ${\bf k}$-point grid
\cite{Wierzbowska2006}, before $\lambda_{\text{EPC}}$ and
$\alpha^2 F(\omega)$
were evaluated using Eqs.~\eqref{eq:lambda} and \eqref{eq:a2f}.
A Gaussian broadening of 0.121~THz was used to approximate
the phonon delta-function
$\delta\left(\omega - \omega_{{\bf q}\nu}\right)$,
and a Gaussian broadening of 0.02~Ry was used to approximate
the electronic delta-functions
$\delta\left(\varepsilon_{m, {\bf k}+{\bf q}} -
\varepsilon_{\text{F}}\right)$
and $\delta\left(\varepsilon_{n, {\bf k}} -
\varepsilon_{\text{F}}\right)$.
The critical temperature $T_{\text{c}}$,
assuming a conventional superconducting mechanism,
was estimated from the calculated $\lambda_{\text{EPC}}$, $\alpha^2 F(\omega)$,
and phonon energies using the modified McMillan formula
\cite{McMillan1968, Dynes1972}:

\begin{equation}
T_{\text{c}} =
\frac{\omega_{\ln}}{1.2}
\exp
\left[
-
\frac{1.04 (1 + \lambda_{\text{EPC}})}
{\lambda_{\text{EPC}} - \mu^\ast (1 + 0.62 \lambda_{\text{EPC}})}
\right],
\label{eq:mcmillan}
\end{equation}
where the logarithmic average phonon energy is defined as
\begin{equation}
\omega_{\ln} =
e^{\frac{2}{\lambda_{\text{EPC}}} \int \frac{d \omega}{\omega}
\alpha^2 F(\omega) \log \omega}.
\label{eq:omegalog}
\end{equation}
In Eq.~\eqref{eq:mcmillan}, a standard value of $\mu^\ast = 0.13$ was used
for the Coulomb pseudopotential.

\begin{figure}[htbp]
\includegraphics[width=1\textwidth]{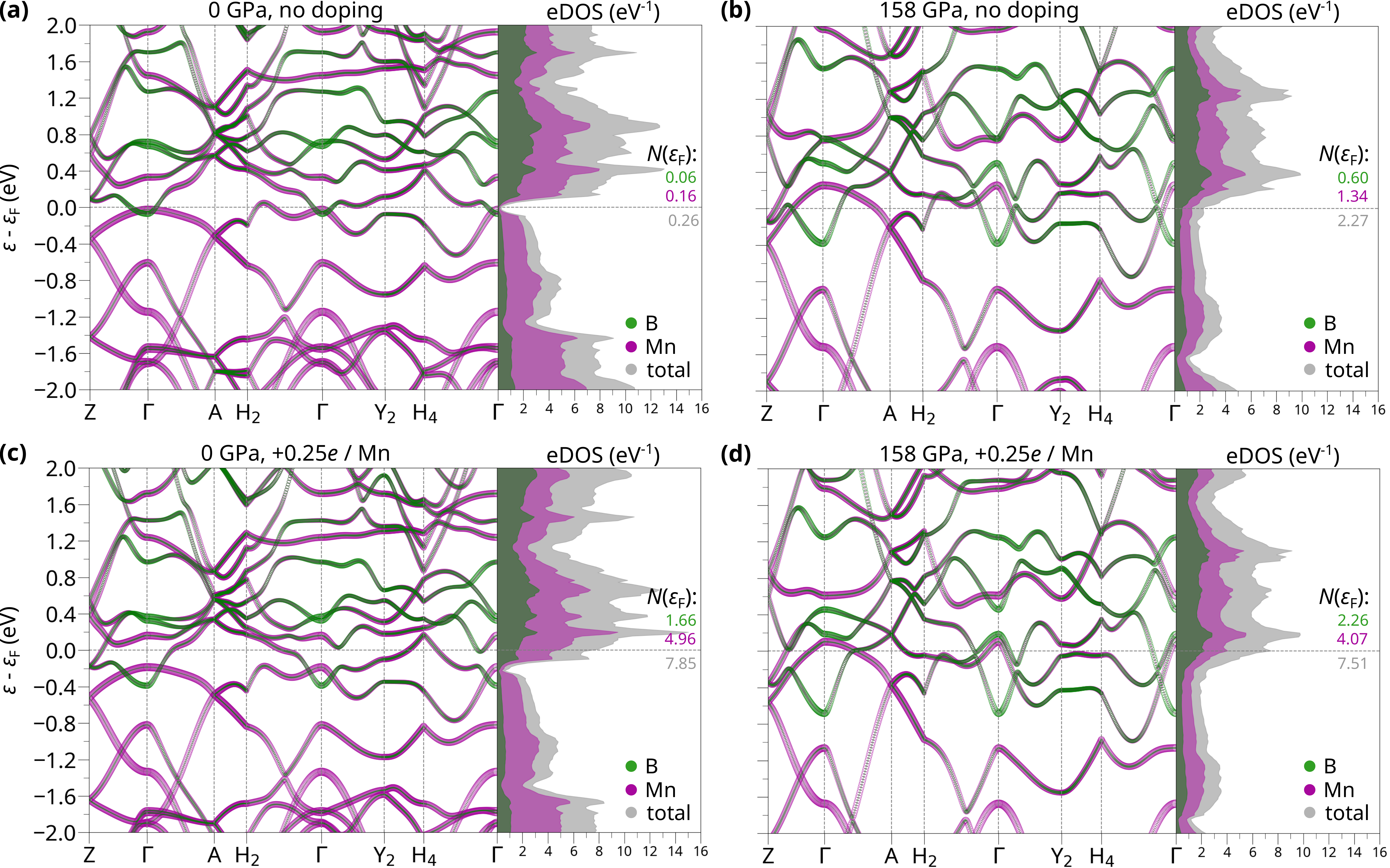}
\caption{Projected electronic band structure and
density of states (eDOS) of
nonmagnetic $P2_1/c$ MnB$_4$ at
  (a) 0 GPa without doping,
  (b) 158 GPa without doping,
  (c) 0 GPa with $+0.25e / \text{Mn}$ doping,
  and
  (d) 158 GPa with $+0.25e / \text{Mn}$ doping.
The eDOS is shown for both spin channels,
with the corresponding values at the Fermi level
$N(\varepsilon_{\text{F}})$
indicated near the horizontal dashed gray line.
}
\label{fig:electrons_doping}
\end{figure}

\begin{figure}[htbp]
  \includegraphics[width=0.9\textwidth]{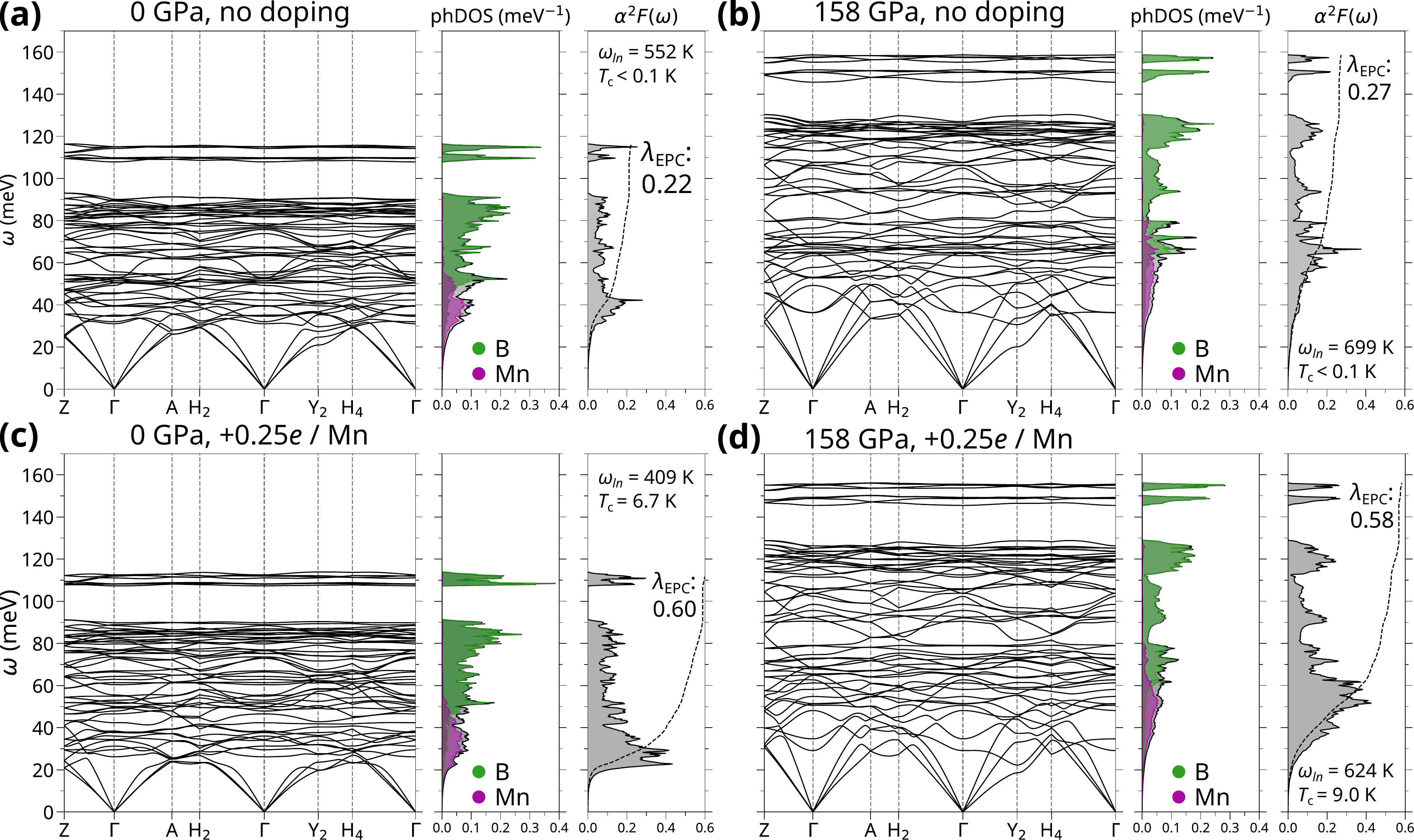}
  \caption{Phonon modes, phonon density of states (phDOS),
  Eliashberg spectral function ($\alpha^2 F(\omega)$), and
  EPC constant ($\lambda_{\text{EPC}}$) of
  nonmagnetic $P2_1/c$ MnB$_4$ at
  (a) 0 GPa without doping,
  (b) 158 GPa without doping,
  (c) 0 GPa with $+0.25e / \text{Mn}$ doping,
  and
  (d) 158 GPa with $+0.25e / \text{Mn}$ doping.
  Mn phDOS: purple;
  B phDOS: green;
  integrated $\alpha^2 F(\omega)$,
  $\int_0^{\omega} \frac{d \omega^\prime}{\omega^\prime}
  2 \alpha^2 F(\omega^\prime)$: dashed black line.
  }
  \label{fig:phonons}
\end{figure}

In this section, the EPC constant
of MnB$_4$
is calculated at 0 and 158~GPa to show that the conventional
mechanism cannot explain the experimental data.
Although there is no \textit{a priori} justification to assume
that MnB$_4$ could be doped, the case
with $1/4$ added electrons per Mn atom is also considered,
as it demonstrates what happens when electronic density of states (eDOS)
at the Fermi level
is increased at 0 and 158~GPa [see
Figs.~\ref{fig:electrons_doping}(a)--\ref{fig:electrons_doping}(d)].
Calculated phonon modes, phonon density of states (phDOS),
$\alpha^2F(\omega)$, $\omega_{\ln}$, $\lambda_{\text{EPC}}$, and $T_{\text{c}}$
are shown with and without doping
at 0 and 158~GPa in Figs.~\ref{fig:phonons}(a)--\ref{fig:phonons}(d).

\begin{figure}[tb]
\includegraphics[width=0.7\columnwidth]{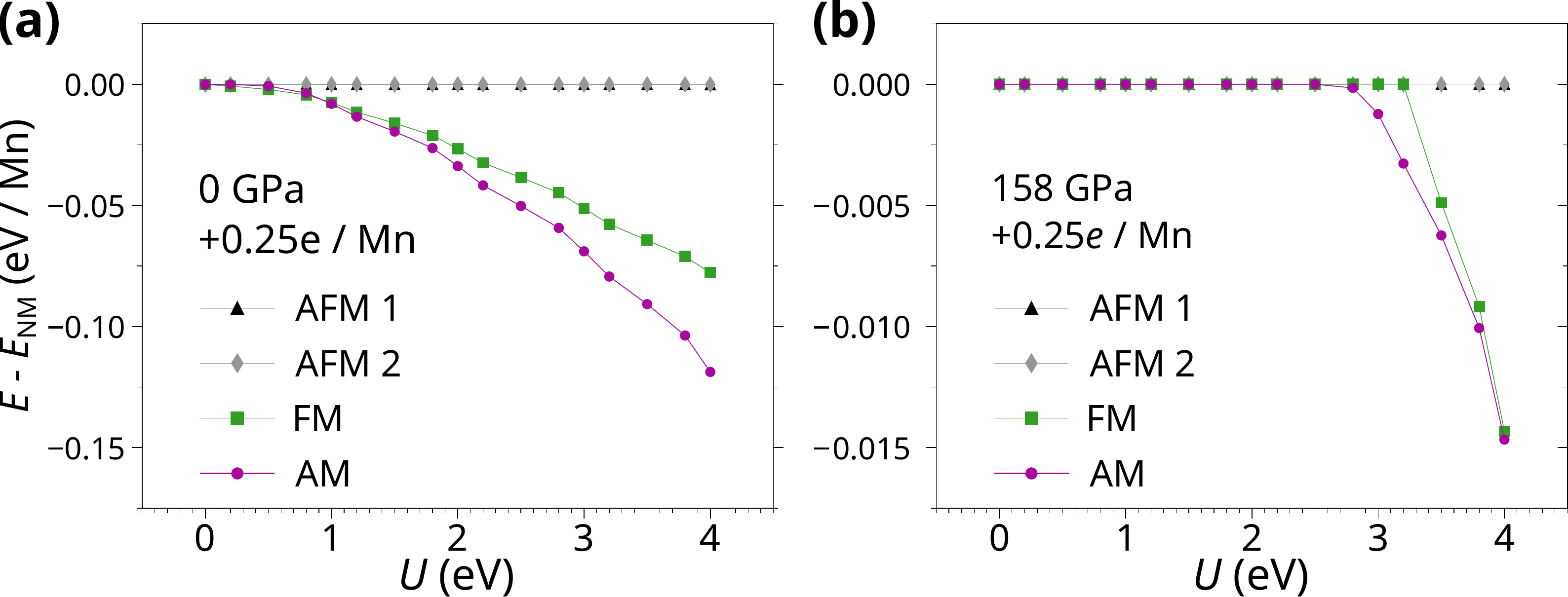}
\caption{Relative energy of
  magnetic
  $P2_1/c$ MnB$_4$ configurations with $+$0.25$e$ / Mn doping
  with respect to
  the nonmagnetic (NM) state as a
  function of the Hubbard U parameter at (a) 0~GPa and (b)
  158~GPa.
  Abbreviations denote antiferromagnetic (AFM 1
  and AFM 2), ferromagnetic (FM), and altermagnetic (AM)
  orders. Note the different energy scales used for 0 and
  158~GPa.}
\label{fig:u_doping}
\end{figure}

In Figs.~\ref{fig:phonons}(a)-\ref{fig:phonons}(b),
the phonon modes without doping
at 0~GPa are softer than those at 158~GPa.
$\alpha^2F(\omega)$ has the highest values in the region dominated
by Mn vibrations.
The resulting $\omega_{\ln}$ values at 0 and 158~GPa are 552~K and 699~K,
respectively, while the corresponding $\lambda_{\text{EPC}}$ values
are 0.22 and 0.27.
$T_{\text{c}}$ estimated with Eq.~\eqref{eq:mcmillan} is below 0.1~K
at both pressures,
in contradiction with the experimental $T_{\text{c}}$ of 14.2~K at 158~GPa
\cite{Xiang2024}.
Notably, these estimates of the EPC constant and critical temperature
are even lower than those reported previously \cite{Xiang2024},
most likely due to a different choice of smearing values
in the approximation of the Dirac delta-functions,
a denser ${\bf q}$-point grid,
and the slightly higher Coulomb pseudopotential used in this work.

Calculated results with the doping at both pressures are shown in
Figs.~\ref{fig:phonons}(c)--\ref{fig:phonons}(d).
The increased eDOS in
Figs.~\ref{fig:electrons_doping}(c)--\ref{fig:electrons_doping}(d)
compared to
Figs.~\ref{fig:electrons_doping}(a)--\ref{fig:electrons_doping}(b)
leads to elevated $T_\text{c}$ estimates.
Although the calculated $T_\text{c}$ of 9~K at 158~GPa
with doping
is closer to the experimental data
than the undoped case,
the doped structure also yields a non-zero $T_\text{c}$ at zero pressure,
which is ruled out by experimental evidence.
These results eliminate the possibility of conventional
superconductivity being driven by doping.

As a side note, in the doped structure,
the altermagnetic (AM) spin fluctuations would compete
with ferromagnetic (FM) fluctuations,
as demonstrated by the DFT$+U$ calculations in
Figs.~\ref{fig:u_doping}(a)--\ref{fig:u_doping}(b).
At 158~GPa, both AM and FM phases would be almost equally likely.

\section{Wannierization \& Tight-binding parameters}
\begin{figure}[htbp]
  \includegraphics[width=1\textwidth]{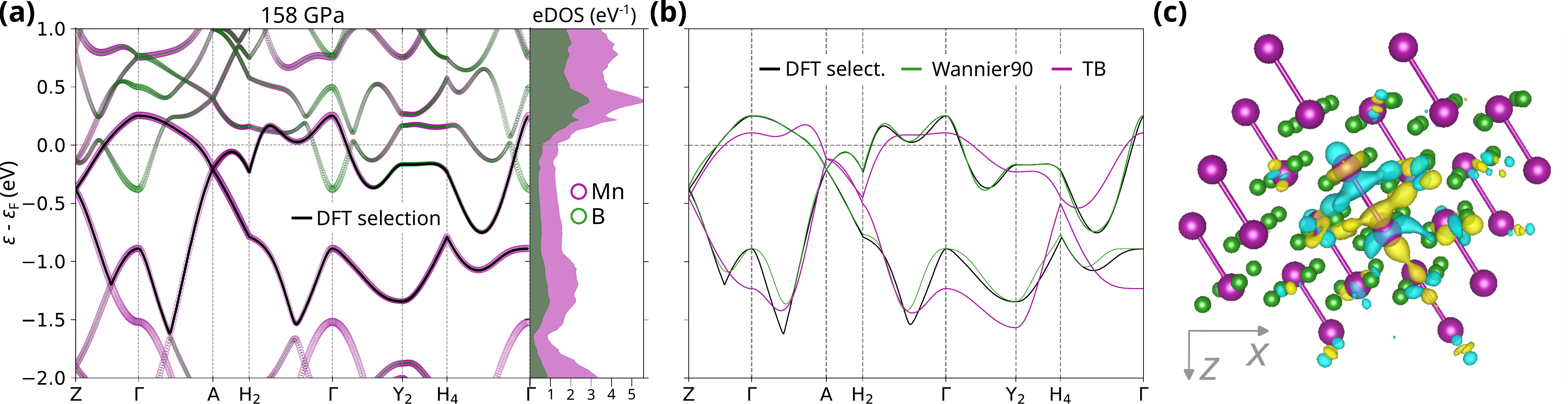}
  \caption{
  Extraction of tight-binding parameters from Mn bands:
  (a) Selection of two DFT bands near the Fermi level
  with dominant Mn character,
  (b) Wannierization and tight-binding fit,
  and
  (c) spatial distribution of one of the two Wannier orbitals
  in a top view.
  In (a), projected Mn and B contributions are indicated in
  purple and green, respectively, with selected DFT bands
  shown in black.
  In (b), selected DFT bands are black, the Wannier90 fit
  is green, and the tight-binding (TB) fit extracted from
  Wannierization is purple.
  In (c), the spatial distribution of the Wannier
  orbital is shown in blue and yellow,
  with gray arrows representing Cartesian axes.
  }
  \label{fig:wan}
\end{figure}

The interface between VASP 6.5.1 and
Wannier90 3.1.0 \cite{Pizzi2020, Marzari2012}
was used for Wannierization.
Two Wannier orbitals were utilized, constrained to the midpoint
of the two inversion centers [blue crosses in Figs.~1(a)--1(d)].
Two bands near the Fermi level with dominant Mn character, as deduced
from the VASP projected bands in Fig.~\ref{fig:electrons_doping}(b),
were fitted [see Fig.~\ref{fig:wan}(a)].
These bands were entangled \cite{Souza2001} with other bands, including
the states with predominantly B content,
precluding the use of a frozen window.
An outer energy window spanning from $-$1.63 to $+$0.26~eV relative
to the Fermi level was employed.

The tight-binding parameters were extracted from
the matrix elements of the 2$\times$2 Hamiltonian
in Wannier space $H_{mn}({\bf R}_p)$ that were
Fourier-interpolated to reciprocal space, $H_{mn}({\bf k})$,
using the atomic gauge notation:
\begin{equation}
\begin{aligned}
H_{mn}({\bf k}) &=
\sum_p^{N_{\text{cells}}}
e^{-i {\bf k} \cdot \left({\bf R}_p + \Delta {\bf r}_{mn}\right)}
H_{mn}({\bf R}_p).
\end{aligned}
\label{eq:ham_atom_gauge}
\end{equation}
In Eq.~\eqref{eq:ham_atom_gauge}, ${\bf R}_p$ are grid vectors identifying
the coordinates of each of the $N_{\text{cells}}$ unit cells $p$
in a supercell,
and $\Delta {\bf r}_{mn} = {\bf r}_m - {\bf r}_n$
are relative vectors between orbitals $n$ and $m$.

The resulting Wannier bands are shown in Fig.~\ref{fig:wan}(b)
in green. The spatial distribution of one of the
corresponding Wannier orbitals is shown in
Fig.~\ref{fig:wan}(c). Hopping parameters for the
tight-binding Hamiltonian were extracted from the Wannier
Hamiltonian restricted to 3 shells of cells. When all
higher-order neighbor terms are excluded, the resulting
tight-binding band structure is represented by purple solid
lines in Fig.~\ref{fig:wan}(b). Although the purple
tight-binding bands exhibit some deviations from the black
DFT and green Wannier bands, this reduction eliminates the
boron degrees of freedom and provides important insight into
the leading interactions within the structure. The
tight-binding fit presented in Fig.~\ref{fig:wan}(b) is the
best achieved in this work. Nonetheless, deviations of the
tight-binding parameters between various fits did not alter
the conclusions regarding the leading pairing channels
reported in this work.

\section{History of MnB$_4$ phase exploration}

Manganese tetraboride (MnB$_4$), synthesized at high temperatures
and brought to ambient conditions,
was experimentally discovered
in two monoclinic structures.
Original powder X-ray diffraction (XRD) measurements
in the late 1960s
indicated the presence of the
$C2/m$ phase [space group (SG) No. 12]
\cite{Fruchart1960, Andersson1969,
Andersson1970, Gou2012}.
However, this structure
has not received further experimental confirmation
in subsequent studies.
Instead, a $P2_1/c$ phase
forming a hard (although brittle) material
was repeatedly confirmed by
XRD
\cite{Knappschneider2014, Gou2014, Bykova2015},
$^{11}$B magic-angle spinning
nuclear magnetic resonance (MAS-NMR) \cite{Knappschneider2014},
and
scanning and
transmission electron microscopy (SEM and TEM) measurements
\cite{Gou2014, Bykova2015}. This semimetallic non-magnetic (NM) phase
was identified as the lowest energy among all
possible candidates by density-functional theory (DFT) calculations
\cite{Wang2011, Niu2012, VanDerGeest2014, Hajinazar2021}.
It was suggested to originate
from an orthorhombic metallic
$Pnnm$ phase  \cite{Knappschneider2014, Gou2014},
which is dominant
among the $M$B$_4$ (with $M = $~Cr, Mn, Fe, Tc, Ru)
materials \cite{Kolmogorov2010, Bialon2011, Niu2012, Gou2013},
due to the Peierls distortion \cite{Kumagai2012}
of the straight Mn chains.
This distortion is marked by the
the introduction of
two alternating Mn--Mn distances
within each chain \cite{Liang2015}.
The longer links of these chains, forming Mn--Mn dimers, are shown
in Figs.~1(a)--1(d).

Further \textit{ab initio} calculations \cite{Liang2015} and
high-temperature powder
XRD measurements \cite{Knappschneider2015} concluded that
the NM $P2_1/c$ phase at ambient conditions
can be reversibly transformed to the ferromagnetic (FM)
or antiferromagnetic (AFM) $Pnnm$ phase at higher
temperatures---approximately 650~K.
Originally, magnetic susceptibility measurements at ambient conditions revealed
weak ferromagnetic spin correlations but no
long-range magnetic order in the synthesized MnB$_4$ samples,
suggesting the material was
paramagnetic \cite{Knappschneider2014}.
Subsequently, the existence of intrinsic magnetism in $P2_1/c$ MnB$_4$ was challenged
by other magnetic susceptibility measurements \cite{Steinki2017},
which concluded
that the weak ferromagnetic/paramagnetic signal
could originate from
Mn impurities
in the samples rather than
from the bulk MnB$_4$ structure.

\bibliography{references}